\documentstyle[12pt]{article}
\input epsf

\begin{document}

\begin{titlepage}

\begin{flushright}
UCLA-00-TEP-08\\
LAPTH  784-00\\
February 2000
\end{flushright}
\vspace{2.5cm}

\begin{center}
\large\bf
{\LARGE\bf Muon colliders and the non-perturbative dynamics of
           the Higgs boson\footnote{Invited talk presented by A. Ghinculov 
           at the 5$^{th}$ International Conference on Physics Potential 
           and Development of $\mu^+\mu^-$ Colliders, 
           December 15-17, 1999, Fairmont Hotel, San Francisco, CA, USA.}}\\[2cm]
\rm
{Adrian Ghinculov$^{a}$ and Thomas Binoth$^b$}\\[.5cm]

{\em $^a$Department of Physics and Astronomy, UCLA,}\\
      {\em Los Angeles, California 90095-1547, USA}\\[.1cm]
{\em $^b$Laboratoire d'Annecy-Le-Vieux de Physique 
         Th\'eorique\footnote{UMR 5108 du CNRS, associ\'ee \`a l'Universit\'e 
         de Savoie.} LAPTH,}\\
      {\em Chemin de Bellevue, B.P. 110, F-74941, 
           Annecy-le-Vieux, France}\\[3.cm]

\end{center}
\normalsize

\begin{abstract}
A muon collider operating in the TeV energy range can be an ideal
$s$-channel Higgs boson factory. This is especially true for a very heavy
Higgs boson. The non-perturbative dynamical aspects of such a
Higgs boson were recently investigated with large $N$ expansion methods
at next to leading order, and reveal the existence of a mass saturation
effect. Even at strong coupling, the Higgs resonance remains always below
1 TeV. However, if the coupling is strong enough, the resonance becomes
impossible to
be detected.
\end{abstract}


\end{titlepage}


\title{Muon colliders and the non-perturbative dynamics of
           the Higgs boson\footnote{Invited talk presented by A. Ghinculov 
           at the 5$^{th}$ International Conference on Physics Potential 
           and Development of $\mu^+\mu^-$ Colliders, 
           December 15-17, 1999, Fairmont Hotel, San Francisco, CA, USA.}}

\author{Adrian Ghinculov$^{a}$ and Thomas Binoth$^b$}

\date{{\em $^a$Department of Physics and Astronomy, UCLA,}\\
      {\em Los Angeles, California 90095-1547, USA}\\
      {\em $^b$Laboratoire d'Annecy-Le-Vieux de Physique 
         Th\'eorique\thanks{UMR 5108 du CNRS, associ\'ee \`a l'Universit\'e 
         de Savoie.} LAPTH,}\\
      {\em Chemin de Bellevue, B.P. 110, F-74941, 
           Annecy-le-Vieux, France}}

\maketitle

\begin{abstract}
A muon collider operating in the TeV energy range can be an ideal
$s$-channel Higgs boson factory. This is especially true for a very heavy
Higgs boson. The non-perturbative dynamical aspects of such a
Higgs boson were recently investigated with large $N$ expansion methods
at next to leading order, and reveal the existence of a mass saturation
effect. Even at strong coupling, the Higgs resonance remains always below
1 TeV. However, if the coupling is strong enough, the resonance becomes
impossible to
be detected.
\end{abstract}


\vspace{.7cm}

A central question in today's particle physics is  
how the electroweak symmetry breaking is realized in nature.
Further experimental
input is needed for distinguishing between various theoretical 
possibilities, and this will be the main goal of the LHC.
The simplest of these possibilities is the minimal scalar sector of 
the standard model which predicts the existence of one single Higgs particle.

The sensitivity of low energy quantum corrections to the mass of the 
Higgs boson is small because of Veltman's screening theorem. Therefore
the indirect Higgs mass determination from radiative corrections is rather
imprecise, in spite of the impressive accuracy of LEP, SLC, and Tevatron 
measurements. Current electroweak data fits based on the minimal 
standard model favor a lighter Higgs boson, with a central value
around 110 GeV, which is close to the region excluded by direct 
production bounds.

So far no significant deviations from the standard model radiative corrections 
were measured which would hint towards the existence of additional degrees of
freedom at higher energy. However, their existence is strongly supported
by well-known open questions of the standard model on the theoretical side.
Such degrees of freedom have the potential to induce additional 
radiative corrections and thus shift the prediction for the Higgs boson
mass. It is conceivable that, once built, the LHC will discover a Higgs
resonance considerably heavier than the central values suggested by 
electroweak data fits at present.

An interesting feature of a possible muon collider is that it can be used as
an $s$-channel Higgs factory. Here we would like to discuss the implications
of the non-perturbative dynamics of the scalar sector for $\mu^+\mu^-$ Higgs 
factories.
We will argue that due to the non-perturbative
dynamics of the scalar sector, a possible muon collider will not need an energy
much higher than 1 TeV to study even a strongly coupled standard Higgs boson.
However, it may need a high luminosity.

A heavy Higgs boson implies a strongly self-interacting scalar sector. Thus
it complicates the theoretical analysis because at some point perturbation 
theory becomes unreliable. A few radiative corrections induced by heavy Higgs
bosons are available in higher order \cite{calc2loop}. 
Their convergence properties were 
studied by several authors \cite{scheme}, and revealed rather 
large theoretical uncertainties. 
In order to avoid the problems of perturbation theory at strong coupling,
such as large renormalization scheme uncertainties and the blow-up of 
radiative corrections in higher loop order, a non-perturbative approach
is necessary.

We performed a study of the
Higgs sector at strong coupling by using non-perturbative $1/N$ expansion
techniques at higher order. This study revealed the existence of an 
interesting mass saturation effect. When the coupling constant 
of the scalar sector is increased, the mass of the Higgs boson 
remains bounded under a saturation value just under 1 TeV, while its 
widths continues to increase. 

Along the lines of 't Hooft's work on planar QCD \cite{thooft:QCD}, the large 
$N$ expansion has attracted a lot of attention by holding the 
promise to solve nonabelian gauge theories non-perturbatively. 
It was also used in the study of critical phenomena.
Its connections to matrix models, two-dimensional gravity, and 
string theory were also explored.

Given that the standard model's Higgs sector is a gauged $SU(2)$
sigma model, the $1/N$ expansion suggests itself naturally for studying
it at strong coupling. At leading order in $1/N$, this was initiated in 
ref. \cite{coleman}. Unfortunately, the leading order solution proves to be 
quite a poor approximation, which in the weak coupling limit
deviates substantially from perturbation theory. Because of this it cannot
be used in realistic phenomenological studies. In ref. \cite{1ovn} we 
extended this study to next-to-leading order. It turns out that 
the next-to-leading order solution is impressively accurate. In the
weak coupling limit it competes with the best perturbative results
available at two-loop precision.

The starting point of the $1/N$ analysis is the Lagrangean of the 
standard model's scalar sector promoted to a $O(N)$-symmetric sigma model.
The well-known equivalence theorem provides a relation between the physics 
of the purely scalar sector and the physics of electroweak vector bosons.
The standard model case is recovered in the $N=4$ limit:

\begin{figure}[t]
\hspace{1cm}
  \epsfxsize = 13cm \epsffile{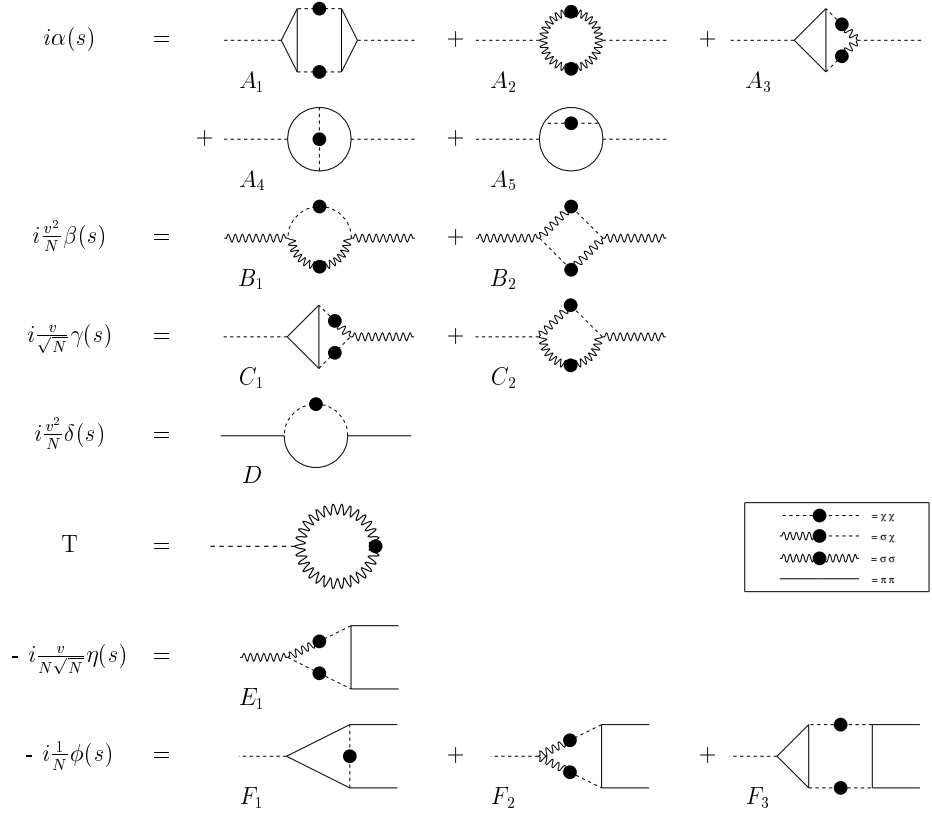}
\caption{{\em Multiloop diagrams which contribute
              in next-to-leading order in $1/N$
              to the two- and three-point functions of the $O(N)$ sigma model.
	      The blobs on propagators denote chains of one-loop 
              Goldstone boson bubble diagrams.}}
\end{figure}

\begin{equation}
  {\cal L}_1 =   \frac{1}{2}           
               \partial_{\nu}\Phi_0 \partial^{\nu}\Phi_0 
             - \frac{\mu_0^2}{2}      \Phi_0^2 
	     - \frac{\lambda_0}{4! N} \Phi_0^4 
 ~~ , ~~ \Phi_0 \equiv \left( \phi_0^1, \phi_0^2, \dots , \phi_0^N \right)
\end{equation}

The next step is to introduce an additional unphysical field $\chi$ in this
Lagrangian \cite{coleman}: 

\begin{eqnarray}
  {\cal L}_2 & = & {\cal L}_1 + \frac {3 N}{2 \lambda_0} 
                 (\chi_0 - \frac{\lambda_0}{6 N} \Phi_0^2 - \mu_0^2)^2 
	\nonumber \\	
           & = &				     
    \frac{1}{2} \partial_{\nu}\Phi_0 \partial^{\nu}\Phi_0 
  - \frac{1}{2} \chi_0 \Phi_0^2 
  + \frac{3 N}{2 \lambda_0} \chi_0^2
  - \frac{3 \mu_0^2 N}{\lambda_0} \chi_0 + const. 
\end{eqnarray}

\begin{figure}[th]
\hspace{1cm}
  \epsfxsize = 13cm \epsffile{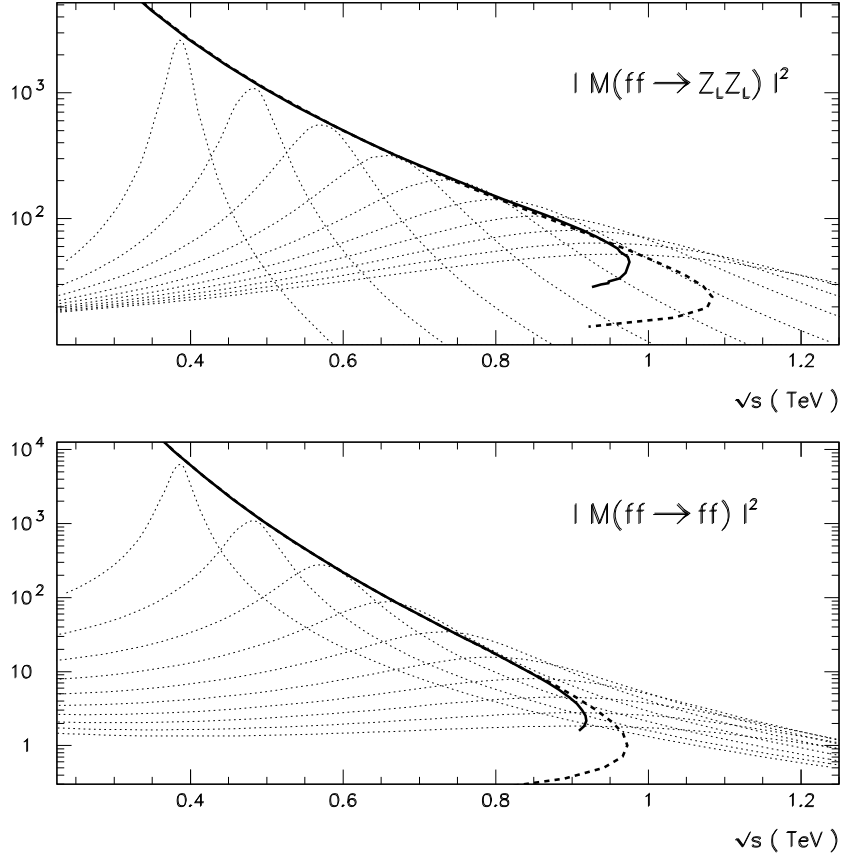}
\caption{{\em The Higgs line shape at a $\mu^+\mu^-$ Higgs factory.
              We marked the position of the maxima of the resonances
              (solid line for the $1/N$ result and dashed line for the
               perturbative result at two-loop).}}
\end{figure}

The auxiliary field $\chi$ does not correspond to
a dynamical degree of freedom. Its equation of motion is simply a constraint
and can be used for eliminating $\chi$. 

While the introduction of
the auxiliary field does not change the dynamics, it does alter the Feynman 
rules by eliminating the scalar quartic couplings. This proves to be extremely
helpful for calculations beyond leading order in $1/N$. Denoting the Higgs
boson by $\sigma$ and the Goldstone bosons by $\pi$, the Feynman rules 
derived from the Lagrangean ${\cal L}_2$ have only trilinear vertices of the
type $\chi\sigma\sigma$ and $\chi\pi\pi$. One can easily count the powers of
$N$ of a Feynman graph by noticing that closed Goldstone loops give rise to
a factor $N$, $\chi\chi$ propagators have a factor $1/N$, and mixed 
$\chi\sigma$ propagators have a factor $1/\sqrt{N}$.

In figure 1 we show the Feynman diagrams which we need for calculating
Higgs production and decay processes at muon colliders at next-to-leading
order in the $1/N$ expansion. These are all one-, two-, and three-point
functions of the sigma model. We note that the summation of leading
order renormalon chains on internal propagators of these diagrams leads 
to an additional Euclidean pole in the propagators. The origin and physical 
content of the tachyon pole is discussed in ref. \cite{1ovn}. 
When effectively calculating the diagrams in figure 1, we use the minimal
tachyonic subtraction discussed in ref. \cite{1ovn} for treating it.

We calculated the diagrams shown in figure 1 numerically, along the lines
of ref. \cite{3loop}. Once they are available numerically, they can be used for
deriving amplitudes of physical processes. Two Higgs processes are
of interest at $\mu^+\mu^-$ $s$-channel Higgs factories: 
$\mu^+\mu^- \rightarrow H \rightarrow t \bar t$ and
$\mu^+\mu^- \rightarrow H \rightarrow ZZ,W^+W^-$.
Their amplitudes are given in the $1/N$ expansion by the following
expression at next-to-leading order:

\begin{eqnarray}
{\cal M}_{f\bar{f}} & = &
\frac{1}{s - m^2(s) \left[ 1 - \frac{1}{N} f_1(s)  \right] }
  \nonumber  \\
{\cal M}_{WW} & = &
\frac{m^2(s)}{\sqrt{N} v}
\frac{1 - \frac{1}{N} f_2 }{s - m^2(s) \left[ 1 - \frac{1}{N} f_1(s)  \right] }
\end{eqnarray}

Here, the correction functions $f_1$ and $f_2$ are given by a combination
of the two- and three-point functions defined in figure 1:

\begin{eqnarray}
f_1(s)  & = &
           \frac{m^2(s)}{v^2} \, \hat{\alpha}(s)
       + 2 \hat{\gamma}(s)
       +   \frac{v^2}{m^2(s)} \left[ \hat{\beta}(s)
                               - 2 \frac{s-m^2(s)}{v^2} \left( \delta Z_{\sigma} - \delta Z_{\pi} \right)
                              \right]
  \nonumber  \\
f_2(s)  & = &
           \frac{m^2(s)}{v^2} \, \hat{\alpha}(s) 
       +   \hat{\gamma}(s)
       -   \hat{\phi}(s)
       -   \frac{v^2}{m^2(s)} \hat{\eta}(s)
\end{eqnarray}

The wave function renormalizations $\delta Z_{\sigma}$, $\delta Z_{\pi}$ 
can be extracted from $\hat \beta$, $\hat \gamma$.  
The hat in the expressions above means that the multi-loop diagrams
are subtracted recursively in the ultraviolet, according to the 
Bogoliubov-Parasiuk-Hepp-Zimmermann procedure \cite{1ovn}.
We note that by performing these ultraviolet subtractions we introduce
a renormalization scale. However, in the final physical correction functions
$f_1$ and $f_2$ this renormalization scheme dependence cancels out.
The final result in manifestly independent of the choice of the 
renormalization scheme.

In figure 2 we show numerical results for the 
$\mu^+\mu^- \rightarrow H \rightarrow t \bar t$ and
$\mu^+\mu^- \rightarrow H \rightarrow ZZ,W^+W^-$ processes of eqs 3.
In both processes the Higgs mass saturation effect shows up. When
the strength of the coupling increases, the peak of the resonance
shifts towards higher energy, up to a saturation value just under
1 TeV, and then starts to shift back towards lower energy. At the
same time, the width continues to increase and the resonance becomes
flat and difficult do detect experimentally.

To conclude, we performed a non-perturbative study of the two main 
Higgs processes of interest at a future muon collider. Due to the
non-perturbative dynamics of the Higgs sector, a standard Higgs particle
is bound to result into a resonance with a peak below 1 TeV. Therefore,
a muon collider will not need energies much larger than 1 TeV to cover
the whole range where a standard Higgs may exist. However, due to 
non-perturbative dynamics, at strong coupling the experimental detection
becomes difficult. 
To measure a flat 
Higgs resonance will require precise knowledge of the backgrounds.
Detection will be a matter of luminosity and not of center of mass
energy.

Finally, if the coupling becomes strong enough, the Higgs boson 
will still remain under 1 TeV but will become impossible to detect
with a given luminosity.

\vspace{.25cm}
{\bf Aknowledgement}
The work of A.G. was supported by the US Department of Energy.
The work of T.B. was supported by the EU Fourth Training
Programme "Training and Mobility of Researchers", Network
"Quantum Chromodynamics and the Deep Structure of Elementary Particles",
contract FMRX-CT98-0194 (DG 12 - MIHT). 



\newpage



\begin{thebibliography}{99}

\bibitem{calc2loop}    A. Ghinculov and J.J. van der Bij,
                      {\em Nucl. Phys. {\bf B436} (1995) 30;}
                      {\em  Nucl. Phys. {\bf B482} (1996) 59;}
		       A. Ghinculov,
                      {\em Phys. Lett. {\bf B337} (1994) 137;
		                   (E) {\bf B346} (1995) 426;}
                      {\em Nucl. Phys. {\bf B455} (1995) 21;}
                      P.N. Maher, L. Durand and K. Riesselmann,
                      {\em Phys. Rev. {\bf D48} (1993) 1061; 
                                  (E) {\bf D52} (1995) 553;}
                       A. Frink, B.A. Kniehl, D. Kreimer, K. Riesselmann,
		      {\em Phys. Rev. {\bf D54} (1996) 4548;} 
                      V. Borodulin and G. Jikia,
                      {\em Phys. Lett. {\bf B391} (1997) 434;}
                      W.J. Marciano and S.S.D. Willenbrock,
                      {\em Phys. Rev. {\bf D37} (1988) 2509;}
                      S. Dawson and S. Willenbrock,
                      {\em Phys. Rev. {\bf D40} (1989) 2880.}

\bibitem{scheme}      B. Kniehl and A. Sirlin,
                      {\em  Phys. Rev. Lett. {\bf 81} (1998) 1373;}
                      {\em  Phys. Lett. {\bf B440} (1998) 136;}
                      T. Binoth and A. Ghinculov,
                      {\em  Phys. Rev. {\bf D56} (1997) 3147;}
                      {\em Phys. Lett. {\bf B394} (1997) 139;}
                      K. Riesselmann,
                      {\em Phys. Rev. {\bf D53} (1996) 6226;}
                      U. Nierste, K. Riesselmann,
                      {\em Phys. Rev. {\bf D53} (1996) 6638;}
                      K. Riesselmann, S. Willenbrock,
                      {\em Phys. Rev. {\bf D55} (1997) 311.}

\bibitem{thooft:QCD}  G. 't Hooft,
                      {\em Nucl. Phys. {\bf B72} (1974) 461;}
                      {\em Nucl. Phys. {\bf B75} (1974) 461;}
                      {\em Phys. Lett. {\bf B119} (1982) 369;}
                      {\em Comm. Math. Phys. {\bf B88} (1983) 1.}

\bibitem{coleman}     S. Coleman, R. Jackiw, H.D. Politzer,
                      {\em Phys. Rev. {\bf D10} (1974) 2491;}
                      H.J. Schnitzer,
                      {\em Phys. Rev. {\bf D10} (1974) 1800;}
                      M.B. Einhorn,
                      {\em Nucl. Phys. {\bf B246} (1984) 75;}
                      R. Casalbuoni, D. Dominici, R. Gatto,
                      {\em Phys. Lett. {\bf B147} (1984) 419.}

\bibitem{1ovn}        T. Binoth, A. Ghinculov,
                      {\em Nucl. Phys. {\bf B550} (1999) 77;}
                      {\em Phys. Lett. {\bf B450} (1999) 220;}
                      {\em Phys. Rev. {\bf D60} (1999) 114003;}
                      A. Ghinculov, T. Binoth and J.J. van der Bij,
                      {\em Phys. Rev. {\bf D57} (1998) 1487;}
                      {\em Phys. Lett. {\bf B417} (1998) 343;}
                      {\em Phys. Lett. {\bf B427} (1998) 343.}
                      
\bibitem{3loop}       A. Ghinculov,
                      {\em Phys. Lett. {\bf B385} (1996) 279.}




\end{thebibliography}
\end{document}